\documentclass{revtex4}

\usepackage{graphicx,epsfig}
\def\TeV{\ifmmode {\,\mathrm{ Te\kern -0.1em V}}\else
                   \textrm{Te\kern -0.1em V}\fi}%
\def\GeV{\ifmmode {\,\mathrm{ Ge\kern -0.1em V}}\else
                   \textrm{Ge\kern -0.1em V}\fi}%
\def\MeV{\ifmmode {\,\mathrm{ Me\kern -0.1em V}}\else
                   \textrm{Me\kern -0.1em V}\fi}%
\def\keV{\ifmmode {\,\mathrm{ ke\kern -0.1em V}}\else
                   \textrm{ke\kern -0.1em V}\fi}%
\def\eV{\ifmmode  {\,\mathrm{ e\kern -0.1em V}}\else
                   \textrm{e\kern -0.1em V}\fi}%

\def\greaterthansquiggle{\raise.3ex\hbox{$>$\kern-.75em\lower1ex
\hbox{$\sim$}}}
\def\lessthansquiggle{\raise.3ex\hbox{$<$\kern-.75em\lower1ex
\hbox{$\sim$}}} 
 \newcommand{\la}{\label}
 \newcommand{\ci}{\cite}
\newcommand{\beqn}{\begin{eqnarray}} \newcommand{\eeqn}{\end{eqnarray}}
\newcommand{\bequ}{\begin{equation}} \newcommand{\eequ}{\end{equation}}
\newcommand{\bsl}{\begin{sloppypar}} \newcommand{\esl}{\end{sloppypar}}

\newcommand{\BC}{\begin{center}}
\newcommand{\EC}{\end{center}}

\begin{document}
\bibliographystyle{revtex}


\title{Impact of beam polarization at a future linear collider}



\author{Gudrid Moortgat--Pick}
\affiliation{Deutsches Elektronen--Synchrotron DESY, Hamburg, Germany}
\email[]{gudrid@mail.desy.de}


\date{\today}

\begin{abstract}
Beam polarization at $e^+e^-$ linear colliders will be a powerful tool for 
high precision analyses. In this paper we summarize the polarization-related 
results for Higgs and electroweak physics, QCD, Supersymmetry 
and alternative theories beyond the Standard Model. 
Most studies were made for a planned linear collider 
operating in the energy range $\sqrt{s}= 500-800$~GeV. In particular we 
work out the advantages of simultaneous polarization of the 
electron and positron beam.
\end{abstract}

\maketitle

\section{Introduction}
\label{sec:intro}
\vspace{-.3cm}
Physics beyond the Standard Model (SM) may well be discovered at 
the run II of Tevatron or at the LHC whose start is planned for 
2006. However, it is well known that a linear collider (LC) will be 
needed for  precise measurements and 
for the detailed exploration of possible New Physics (NP). A 
LC will also make possible measurements of the SM with unprecedented
precision. Moreover the chiral character of 
the couplings can be worked out by using beam polarization. The 
importance of such measurements and the physics accessible with polarized 
electrons has been discussed for example in references \ci{Snowmass96,TDR}.

We will use the
convention that, if the sign is explicitly given, $+$ $(-)$ polarization
corresponds to R (L) chirality with helicity $\lambda=+\frac{1}{2}$
($\lambda=-\frac{1}{2}$) for both electrons and positrons.
In the limit of vanishing electron mass
SM processes in the s--channel are initiated by
electrons and positrons polarized in the same direction, i.e. $e^+_Le^-_R$ (LR)
or $e^+_Re^-_L$ (RL), where the first (second) entries 
denote helicities of corresponding particle.
This result follows from the vector nature of
$\gamma$ or $Z$ couplings (helicity--conservation).
In theories beyond the SM interactions also   
(LL) and (RR) configurations from s-channel contributions are allowed and
the polarization of both beams offers a powerful tool for
analyzing the coupling structure of the process as well as 
for enhancing rates and suppressing SM backgrounds. 
We assume that an electron polarization of $P_{e^-}=\pm 80\%$ 
(denoted by $(80, 0)$)
is reachable \cite{Baltay} with an simultaneous
positron polarization of about
$P_{e^+}=\pm 40\%$ (denoted by $(80, 40)$) with no loss of intensity and about
$P_{e^+}=\pm 60\%$ with 55\% of beam intensity \cite{TDR}.

In this paper we explore the physics consequences of beam polarization, in 
particular when both the
electron and positron beams are polarized \cite{Steiner}.
Most studies were made for a planned linear collider 
operating in the energy range $\sqrt{s}= 500-800$~GeV.
The results show that there are
six principal advantages to be gained when both beams are polarized:
(1) higher effective polarization 
$P_{eff}=(P_{e^-}-P_{e^+})/(1-P_{e^-}P_{e^+})$,  
(2) suppression of background (3)
enhancement of rates (${\cal L}$)
(4) increased sensitivity to non-standard
couplings, (5) test of chiral quantum numbers of SUSY scalar particles, and
(6) improved accuracy in measuring the polarization.
These features will be discussed in greater detail in the following sections.  
In particular both for SUSY and for high precision
studies in electroweak physics the polarization of both beams is crucial.

\section{Higgs Physics}
\vspace{-.3cm}
In order to establish experimentally the Higgs mechanism
as the mechanism of electroweak symmetry breaking 
an accurate study of the production and decay properties of Higgs candidates
is needed. The
study of Higgs particles will therefore represent a central
theme of the physics programme of a future LC.  

Higgs production at a LC occurs mainly via $WW$ fusion, 
$e^+e^-\to H \nu \bar{\nu}$, and Higgsstrahlung, $e^+ e^-\to HZ$. 
Polarizing both beams enhances the signal and suppresses background.
The scaling factors, i.e.\ ratios of
polarized and unpolarized cross section, are given in
Table ~\ref{tab_higgs1} \ci{Desch,Steiner}. 
Beam polarization can help to measure the $HZZ$ and the $HWW$ coupling
separately e.g.~via suppression of the $WW$ background
(and the signal of $WW$ fusion) and enhancement of the $HZ$ contribution with
right polarized electrons and left polarized positrons.
Further, variation of the relative amounts of
Higgs-strahlung and $WW$ fusion makes it possible to keep
the systematics arising from
the  contributions to the fitted spectrum for these two processes 
smaller than the statistical accuracy.

Moreover beam polarization reduces considerably the error when
determining the Higgs couplings.
In an effective Lagrangian approach the general coupling between
Z--, Vector-- and Higgsboson can be written:
\beqn
{\cal L}&=&(1+a_Z)\frac{g_Z m_Z}{2} H Z_{\mu} Z^{\mu} +\frac{g_Z}{m_Z}
\sum_{V=Z,\gamma} \big[ b_V H Z_{\mu\nu}V^{\mu\nu}
+c_V(\partial_{\mu} H Z_{\nu}-\partial_{\nu} H Z_{\mu}) V^{\mu\nu}
+\tilde{b}_V H Z_{\mu\nu} \tilde{V}^{\mu\nu}\big],\la{eq_higgs1}
\eeqn
with $V_{\mu\nu}=\partial_{\mu} V_{\nu}-\partial_{\nu} V_{\mu}$,
$\tilde{V}_{\mu\nu}=\epsilon_{\mu\nu\alpha\beta} V^{\alpha \beta}$.

Using, for example, the optimal--observable method 
it is possible at a LC to determine the seven complex 
Higgs couplings with high accuracy: the CP--even
$a_Z$, $b_Z$, $c_Z$ and $b_{\gamma}$, $c_{\gamma}$ and the CP--odd
$\tilde{b}_Z$ and $\tilde{b}_{\gamma}$. 
Simultaneous beam polarization considerably improves the accuracy. 
A study was made for $\sqrt{s}=500$~GeV and 
${\cal L}=300$~fb$^{-1}$ \ci{Kniehl}. 
It shows that the $ZZ\Phi$ coupling is well
constrained. However, to fix the $Z\gamma \Phi$ coupling  
beam polarization is essential, Table~\ref{tab_higgs2}.
Simultaneous beam polarization $(\pm 80,\mp 60)$
of $e^-$ and $e^+$ beams results in an further reduction 
of 20\%--30\% in the optimal errors
compared to the case $(\pm 80,0)$.

\begin{table}
\parbox{16.5cm}{\caption{Higgs production in Standard Model: 
Scaling factors, i.e. ratios of polarized and unpolarized
cross section $\sigma^{pol}/\sigma^{unpol}$,
are given in Higgs production and background processes 
for different polarization configurations with
$|P_{e^-}|=80\%$, $|P_{e^+}|=60\%$ \ci{Desch,Steiner}. 
\label{tab_higgs1}}}\vspace{.2cm}
\hspace*{-2cm}\begin{tabular}{|l||c|c||c|c|}
\hline
Configuration 
& \multicolumn{2}{|c||}{Higgs Production} &
\multicolumn{2}{||c|}{Background} \\
$(sgn(P_{e^-}) sgn(P_{e^+}))$
&$e^+e^-\to H \nu \bar{\nu}$ & $e^+ e^-\to HZ$ & 
$e^+e^-\to WW$, $e^+e^-\to Z \nu \bar{\nu}$ & $e^+e^-\to ZZ$ \\ \hline
$(R0)$ & 0.20 & 0.87 & 0.20 & 0.76 \\
$(L0)$ & 1.80 & 1.13 & 1.80 & 1.25 \\ \hline
$(RL)$ & 0.08 & 1.26 & 0.10 & 1.05 \\
$(LR)$ & 2.88 & 1.70 & 2.85 & 1.91 \\ \hline
\end{tabular}
\end{table}

\begin{table}
\parbox{17.5cm}{\caption{Determination of
general Higgs couplings: Optimal errors on general $ZZ\Phi$ and 
$Z\gamma \Phi$ couplings for different beam polarizations
\ci{Kniehl}. \label{tab_higgs2} }}\vspace{.2cm}
\hspace*{-2.5cm}\begin{tabular}{|l||c|c|c|}
\hline
 & $P_{e^-}=0=P_{e^+}$ & $P_{e^-}=80\%$, $P_{e^+}=0$ &
$P_{e^-}=80\%$, $P_{e^+}=60\%$ 
\\ \hline
Re($b_Z$) & 0.00055 & 0.00028 & 0.00023 \\
Re($c_Z$) & 0.00065 & 0.00014 & 0.00011 \\ \hline
Re($b_{\gamma}$) & 0.01232 & 0.00052 & 0.00036 \\
Re($c_{\gamma}$) & 0.00542 & 0.00011 & 0.00008 \\ \hline
Re($\tilde{b}_Z$) & 0.00104 & 0.00095 & 0.00078 \\
Re($\tilde{b}_{\gamma}$) & 0.00618 & 0.00145 & 0.00101 \\ \hline
Im($\tilde{b}_Z$) & 0.00521 & 0.00032 & 0.00022 \\
Im($\tilde{b}_{\gamma}$) & 0.00101 & 0.00032 & 0.00026  \\ \hline
\end{tabular}
\end{table}

\section{Electroweak Physics}
\vspace{-.3cm}
At TESLA \cite{TDR} 
it is possible to test the SM with unprecedented accuracy 
\cite{Sngiga}.
At high $\sqrt{s}$ studies determining the triple gauge
couplings \ci{Moenig,Menges} and at low $\sqrt{s}$ 
an order--of--magnitude improvement in the accuracy of
the determination of
$\sin^2\Theta_{eff}^l$ at $\sqrt{s}=m_Z$ may well be possible
\ci{Moenig,Weiglein2}.

{\it GigaZ:} Beam polarization of both $e^-$ and $e^+$ at GigaZ would 
make possible the most sensitive test of the SM ever made
by significantly reducing
the polarization error when using the Blondel Scheme \cite{Blondel}
coupled with Compton polarimetry.
In the SM the left--right asymmetry $A_{LR}$ in the process
$e^+ e^-\to Z\to \ell^+ \ell^-$ depends only on the effective leptonic
mixing. Applying the Blondel Scheme means that $A_{LR}$ is directly
expressed by the cross sections for
the production of Z's with longitudinally polarized beams:
\bequ
\mbox{\hspace{-1cm}}
A_{LR} = \sqrt{\frac{(\sigma^{RR}+\sigma^{RL}-\sigma^{LR}-\sigma^{LL})
(-\sigma^{RR}+\sigma^{RL}-\sigma^{LR}+\sigma^{LL})}
{(\sigma^{RR}+\sigma^{RL}+\sigma^{LR}+\sigma^{LL})(-\sigma^{RR}+\sigma^{RL}
+\sigma^{LR}-\sigma^{LL})}}.\la{eq_ew2b}
\eequ
In this case measurement of the cross
sections for all spin combinations (RR), (RL), (LR), (LL) can be used
to determine the effective polarization
and it is not necessary to
know the beam  polarization with extreme accuracy.
Fig.~\ref{fig_ew1} shows the statistical error on $A_{LR}$ as a function of 
the positron polarization for $P_{e^-}=80\%$.
Already with about 20\% positron polarization the goal
of $\delta \sin^2\theta_{eff} \sim 10^{-5}$ can be reached.
The Blondel scheme also requires some
luminosity for the less favoured combinations (LL) and (RR). However only about
10\% of running time will be needed for these combinations
to reach the desired accuracy for these high precision measurements.
The  Blondel Scheme has the additional advantage
that the polarization measured in this way is the luminosity-weighted value
at the interaction point, rather than the value at the location of the
polarimeter.

{\it High $\sqrt{s}$:}
The production $e^+ e^-\to W^+ W^-$ occurs 
in lowest order via $\gamma$--, $Z$-- and
$\nu_e$--exchange. In order to test the SM with high precision one can 
carefully study triple gauge boson couplings. 
These couplings can be determined by measuring the angular
distribution and polarization of the $W^{\pm}$'s.
Simultaneously fitting of all couplings results in a strong 
correlation between the $\gamma-$ and $Z-$couplings whereas 
polarized beams are well 
suited to separate these couplings.
TESLA with its high luminosity is a very
promising device to measure these couplings with high precision: 
At $\sqrt{s}=500$~GeV and with $|P_{e^-}|=80\%$ statistical
errors of O($10^{-4}$) can be reached.
Moreover, using simultaneous beam polarization $(80,60)$ 
the errors can be further reduced by up to a factor 1.8 
compared to the case with $(80,0)$.
\ci{Menges}. An further advantage of using polarized $e^-$ and $e^+$ beams 
is that one could gain about a factor two 
in running time by using the optimal beam configuration \cite{Moenig}.

\section{QCD physics}
\vspace{-.3cm}
Strong--interaction measurements at a future LC 
will form an important component of 
the physics programme. We restrict ourselves in this section to the study  
of polarization effects as a tool for determining
a) the top couplings and b) polarized $\gamma$ 
structure functions.

{\it Production of tops and FCN couplings:}
High precision measurements of the properties and the interaction
of top quarks will be an essential part of the LC research program
since the top as heaviest known elementary particle probably plays a key 
role in pinning down the origin of electroweak symmetry breaking.
In \ci{Kuehn} polarization effects were studied at the top threshold.
The main background comes from $e^+ e^-\to W^+ W^-$. The scaling
factors for suppressing this background are shown in Table~\ref{tab_higgs1}. 
The gain in using simultaneously polarized $e^-$ and $e^+$ 
beams $(80,60)$ is given by the higher effective 
polarization of $P_{eff}=0.946$ compared to the case for only polarized 
electrons so that the top vector couplings $v_t$ can be measured
up to $1\%$ with ${\cal L}=300$~fb$^{-1}$. 
The advantage of using polarized $e^-$ and $e^+$ beams
has also been studied for deriving limits on
top flavour changing neutral couplings (FCN) 
from single top production and its FCN decays
\ci{Aguilar}. With $e^-$ and $e^+$ polarization (80, 45), 
limits are improved by about a factor 2.5 compared to unpolarized beams, 
wheras in each case the 
positron polarization improves the limits obtained with only electron
polarization by 30\%--40\%.
These improvements correspond to an increase in rate of a factor of 6--7. 

{\it Polarized structure functions} (PSF) {\it of photons:}
For the  LC $\gamma \gamma$, $\gamma e^-$ and $e^- e^-$ modes are
conceivable, and these could be used to
study polarized
structure functions of photons. For TESLA these options are discussed
as a possible upgrade, but it is already possible to get information about
PSF even in the normal $e^+ e^- $ mode if one uses highly polarized $e^+$ and
$e^-$ beams in the process
$e^+ e^-\to \gamma \gamma + e^+ e^-\to \mbox{Di-jets}+e^+e^-$ \ci{Stratmann}.
Since depolarization tends to be
large at the $e \gamma$ vertex one needs highly polarized $e^-$ and $e^+$
beams to get first experimental hints on polarized PSF.

\section{Alternative Theories}
\vspace{-.3cm}
{\it Search for additional gauge bosons $Z'$, $W'$ and for 
contact interactions:}
Beam polarization is a helpful tool to enlarge the discovery reach of $Z'$, 
$W'$ due to higher effective polarization and correspondingly a higher 
luminosity for specific channels, 
but the predicted effects are strongly model dependent.
With $(80,60)$ the discovery reach is increased by 
$10\%$--$20\%$ compared to the case when $(80,0)$ \cite{Riemann}. 
Beam polarization is also important to distinguish between different models of
contact interactions.
Simulation studies are given in \ci{Riemann}. Using $(80,40)$ 
instead of only $(80,0)$
could enlarge the discovery reach for the scale $\Lambda$ of contact 
interactions in $e^+ e^-\to b \bar{b}$ by up to $40\%$ 
for RR or RL interactions. 

{\it Search for large extra dimensions:}
In the direct search for extra dimensions,
$e^+ e^-\to \gamma G$, beam polarization enlarges the discovery reach for 
the scale $M_D$ \ci{Vest}, and is a crucial tool for suppressing
the dominant background $e^+ e^-\to \nu \bar{\nu} \gamma$ \ci{Wilson}.
In the case of two extra dimensions the reach is enlarged by $16\%$
with simultaneous beam polarization
$(80,60)$ compared to the case 
with only electron polarization. Furthermore 
the background can be significantly reduced, the ratio 
$\frac{S}{\sqrt{B}}$ is improved by a factor 2.2 for $(80,0)$ and
by a factor 5 for $(80,60)$.
This corresponds to an increase in rate by a factor 5 compared to when  
only electrons are polarized, and a factor 25 when both beams are polarized.  

\section{SUSY Physics}
\vspace{-.3cm}
Polarization effects play a crucial role in discovering SUSY and in the
determination of supersymmetric model parameters.
Simultaneous polarization of both beams could lead to an 
additional increase of the scaling factor up to an factor 1.6 for realistic 
positron polarizations compared to the case of only polarized electrons, 
depending on the process and on the scenario \cite{Steiner}. 
This enhancement can not be expressed by the 
effective polarization, because these rates depend explicitly on the 
polarization of both beams.
In the following, however, we do not focus on these statistical effects of 
beam polarization but on the determination of the underlying SUSY model. 
In SUSY models all coupling structures consistent with Lorentz invariance
should be considered. Therefore it is possible to get appreciable 
event rates for polarization configurations that are unfavorable for SM 
processes.

All numerical values quoted below, if not otherwise stated, 
are given for the LC--reference scenario for low $\tan\beta$
with the SUSY parameters $M_2=152$~GeV, $\mu=316$~GeV, $\tan\beta=3$ and 
$m_0=100$~GeV \ci{Blair}.

{\it Stop Sector:}
In \ci{Kraml} the feasibility of 
determining the stop mixing angle 
in the process $e^+ e^- \to \tilde{t}_1 \tilde{t}_1$ at TESLA has been 
investigated.
The study was made at $\sqrt{s}=500$~GeV, ${\cal L}=2\times 500$~fb$^{-1}$
and polarization
 $(80,60)$ for the parameters $m_{\tilde{t}_1}=180$~GeV,
$\cos\Theta_{\tilde{t}_1}=0.57$. The resulting errors are
 $\delta(m_{\tilde{t}_1})=1.1$~GeV and
$\delta(\cos\Theta_{\tilde{t}_1})=0.01$. If only polarized electrons  
were used then these errors
would increase by about 20\%.

{\it Slepton sector:}
Beam polarization is a useful tool to improve the accuracy of the end--point 
method for determining the selectron masses \cite{Nauenberg}.
Furthermore with beam polarization the 
association between the chiral fermions and their scalar SUSY partners can be 
established:
$e^-_{L,R} \stackrel{Susy}{\to} \tilde{e}^-_{L,R}$,
$e^+_{L,R} \stackrel{Susy}{\to} \tilde{e}^+_{R,L}$.
The production of sleptons $e^+ e^- \to \tilde{e}_L \tilde{e}_L$,
$e^+ e^- \to \tilde{e}_R \tilde{e}_R$
proceeds via $\gamma$ and $Z$ exchange in the direct channel and
$\tilde{\chi}^0_i$ exchange in the crossed channels and with the intial 
configurations $e^-_L e^+_R$ and $e^-_R e^+_L$.
The mixed production $\tilde{e}_L \tilde{e}_R$ is only possible via the 
crossed channels and with the extraordinary beam configurations:
$e^+_L e^-_L \to \tilde{e}^-_L \tilde{e}^+_R$ and
$e^+_R e^-_R \to \tilde{e}^-_R \tilde{e}^+_L$, and allows
to test the association between chiral leptons with the weak quantum numbers
$R$, $L$ and their scalar partners \cite{slep_01}.
For this test the polarization of both beams
is indispensable since the suppression of the s--channel is not possible with 
only polarized electrons. 

\begin{figure}
\setlength{\unitlength}{1cm}
\begin{center}
\begin{minipage}{7cm}
\hspace*{2.5cm}
\begin{picture}(7,7)
\put(-.2,0){\includegraphics{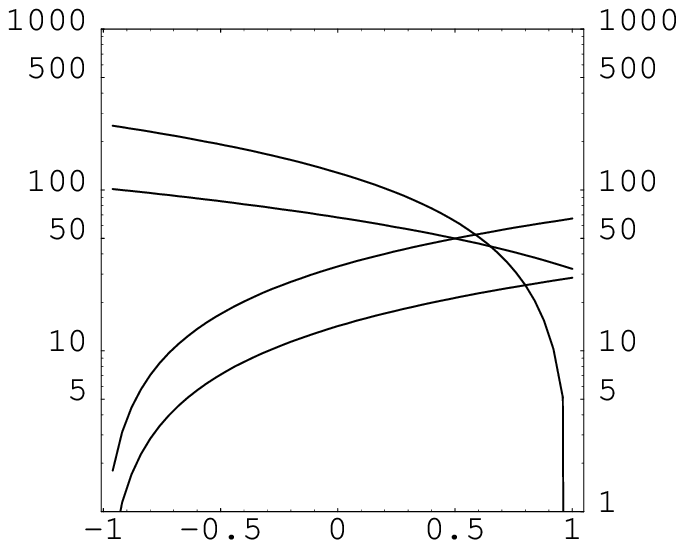}}
\put(5.7,8){b) $e^+ e^-\to \tilde{\chi}^0_1 \tilde{\chi}^0_2$}
\put(-2.5,7.8){\makebox(0,0)[bl]{{
   a) $\sigma(e^+ e^-\to \tilde{e}^+_{R,L} \tilde{e}^-_{R,L}$)/fb}}}
\put(-1.2,6.9){$\tilde{e}^{-}_{L} \tilde{e}^{+}_{R}$}
\put(.7,6.4){$\tilde{e}^{-}_{L} \tilde{e}^+_{L}$}
\put(-.4,4.9){$\tilde{e}^-_{R} \tilde{e}^+_{L}$}
\put(-2.5,5.9){$\tilde{e}^-_{R} \tilde{e}^+_{R}$}
\put(1.7,3.1){$P_{e^+}$}
\end{picture}\par\vspace{.7cm}\hspace{-1cm}
\end{minipage}
\hspace*{-.5cm}
\epsfxsize=5.5cm \epsfbox{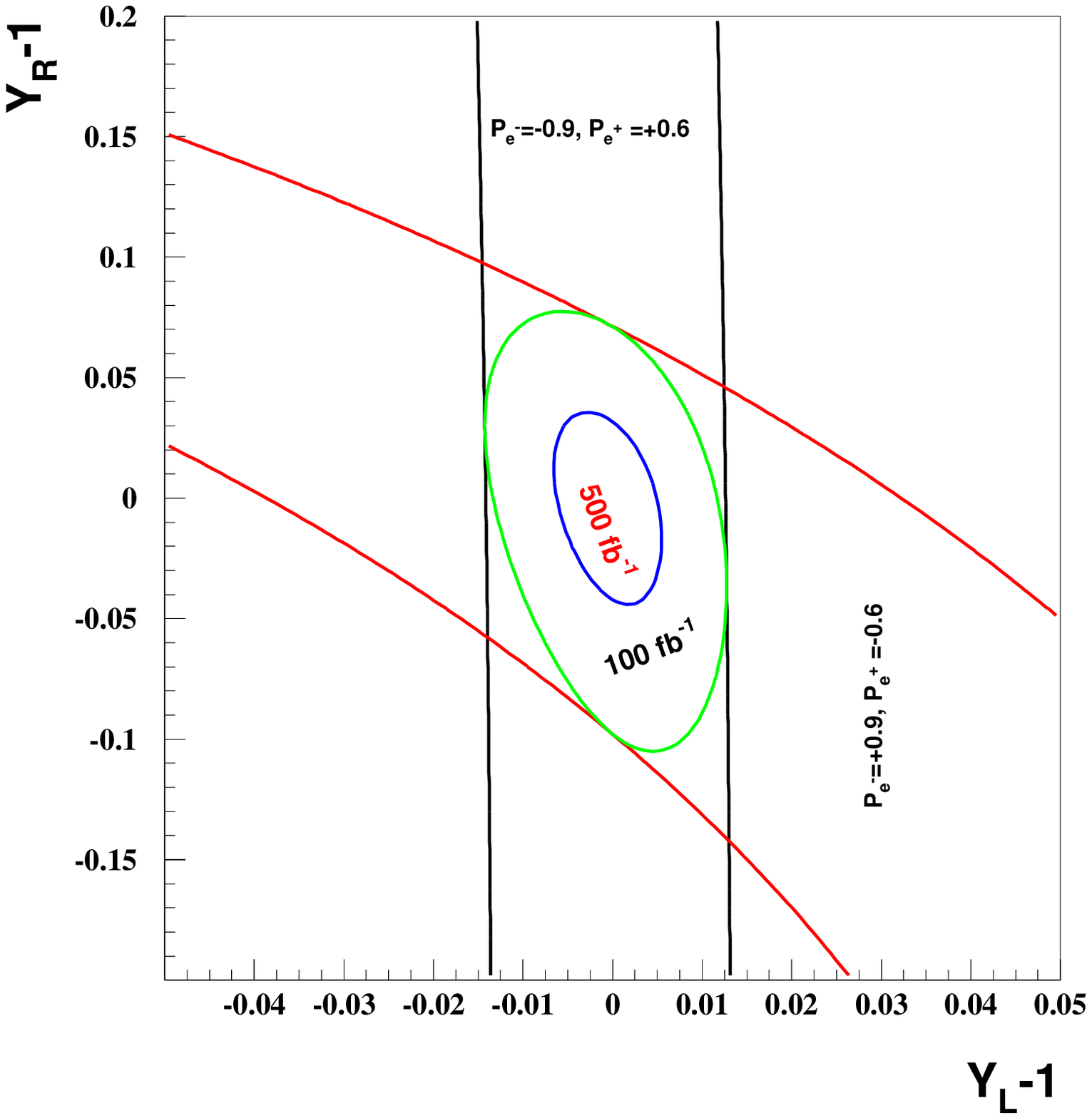}
\vspace*{-2cm}
\end{center}\par\vspace{-2.6cm}
\caption[]{a) Production cross sections as a function of $P_{e^+}$ for
    $\sqrt{s} = 450$~GeV, $P_{e^-}=-0.8$, $m_{\tilde{e}_R}=137.7$ GeV,
  $m_{\tilde{e}_L}=179.3$ GeV, $M_2=156$~GeV, $\mu=316$~GeV and
  $\tan\beta=3$. , $\mu=316$~GeV and $\tan\beta=3$.
  ISR corrections and beam strahlung are included \cite{slep_01};
b) Contours of the cross sections $\sigma_L\{12\}$ and
       $\sigma_R\{12\}$ in the plane of the Yukawa couplings $g_{_{\tilde{W}}}$
       and $g_{_{\!\tilde{B}}}$ normalized to the SU(2) and U(1) gauge
       couplings $g$ and $g'$ $\{Y_L=g_{{\tilde W}}/g,\,
       Y_R=g_{{\tilde B}}/g'\,\}$ for the set ${\sf RP1}$ at the $e^+e^-$
       c.m. energy of 500 GeV; the contours correspond to the integrated
       luminosities 100 and 500 fb$^{-1}$ and the longitudinal polarization
       of electron and positron beams of 90\% and 60\%, respectively.}
\label{fig_slep}
\end{figure}

We show polarized cross sections including ISR and beamstrahlung  for the 
different selectron pair production at $\sqrt{s}=450$~GeV.  
For $P(e^-)=-80\%$ and variable $P(e^+)$ one sees from Fig.~\ref{fig_slep}a
that for $P(e^+)<40\%$ the significantly 
highest rates are those for the pair 
$\tilde{e}^-_L\tilde{e}^+_R$, at least two times larger than for 
all other pairs. 
This clear distinction between the different production
channels is only possible for energies
close to the threshold since for higher energies the effects are covered
by kinematical reasons. 

At an $e^-e^-$ collider slepton production occurs 
via t--channel exchange. It is only possible to verify
the association between $e^-_{L,R}$ and $\tilde{e}^-_{L,R}$.

{\it Chargino sector:} 
In the MSSM the chargino production depends on the fundamental parameters
$M_2$, $\mu$, $\tan\beta$, $m_{\tilde{\nu}_e}$. 
For completely longitudinally polarized beams and assuming 
that the masses of the exchanged
sneutrinos $m_{\tilde{\nu}_e}$ are known,  it has been shown \ci{Choi} 
that these parameters can be determined quite well. 
Furthermore a method has been shown
to constrain $m_{\tilde{\nu}_e}$ indirectly even if the direct production
of $m_{\tilde{\nu}_e}$
is beyond the kinematical reach \ci{Gudi_Char_Neut}, since
the forward--backward--asymmetry of the decay electron in 
$e^+ e^-\to \tilde{\chi}^+_1 \tilde{\chi}^-_1$, 
$\tilde{\chi}^-_1\to \tilde{\chi}^0_1 e^- \bar{\nu}$, is very sensitive to 
$m_{\tilde{\nu}_e}$. 
With additional positron beam polarization one gets further
increase in the rates by a factor of about 1.6, so that 
the statistical error in $\Delta A_{FB}$ is reduced by 20\%. 
 
In single chargino production, $e^+ e^-\to \tilde{e} \tilde{\chi}^- \nu_e$,
$e^+ e^-\to \tilde{e} \tilde{\chi}^+ \bar{\nu}_e$ \ci{Baltay}
the preferred beam polarization configurations 
are (RR) and (LL), which are  
disfavoured in the SM. Since one expects small event rates 
positron polarization could play a major 
role in the measurement and analysis of this process.

{\it Neutralino sector:}
As in the cases studied before,
beam polarization is crucial for a comprehensive 
determination of the 
fundamental parameters, and in particular of $M_1$ \ci{CKMZ}.
Furthermore neutralino production in lowest order occurs via $Z$, 
$\tilde{e}_L$ and $\tilde{e}_R$ exchange and
is sensitive to the chiral couplings and  
the masses of $\tilde{e}_L$, $\tilde{e}_R$. Therefore
the ordering of magnitude 
of the cross sections for different polarization configurations
depends significantly on the character of the neutralinos \ci{Gudi_Char_Neut}.

A linear collider with polarized beams offers even the possibility
to verify very accurately
the fundamental SUSY assumption that the Yukawa couplings,
$g_{_{\tilde{W}}}$ and $g_{_{\!\tilde{B}}}$ are indentical to the SU(2) and
U(1) gauge couplings $g$ and $g'$.
Varying the left--handed and right--handed Yukawa
couplings leads to a significant change in the corresponding left--handed
and right--handed production cross sections. Combining the
measurements of the polarized cross sections
$\sigma_R$ with $(+90,-60)$ and $\sigma_L$ with $(-90,+60)$ for the process
$e^+e^-\rightarrow \tilde{\chi}^0_1\, \tilde{\chi}^0_2$, the
Yukawa couplings $g_{_{\tilde{W}}}$ and $g_{_{\!\tilde{B}}}$ can be
determined to quite a high precision as demonstrated in Fig.~\ref{fig_slep}b.
The $1\sigma$ statistical errors have been derived
for an  integrated luminosity of
$\int {\cal L}\, dt =100$ and $500$ fb$^{-1}$ and for
$P(e^-)=\pm 90\%$, $P(e^+)=\mp 60\%$.

\begin{center}
\begin{figure}
\vspace*{-2cm}
\setlength{\unitlength}{1mm}
\hspace*{-8.5cm}
\begin{picture}(83,73)
\put(0,0){\includegraphics{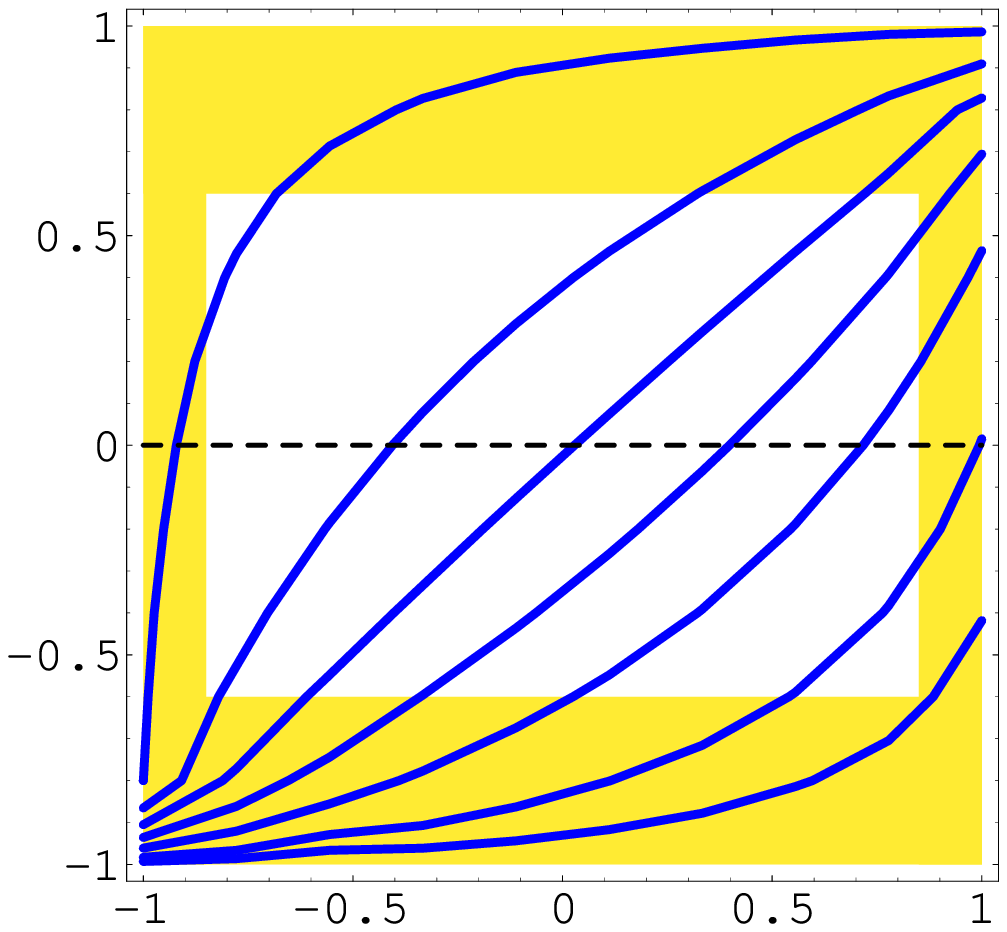}}
\put(20,-100){\includegraphics{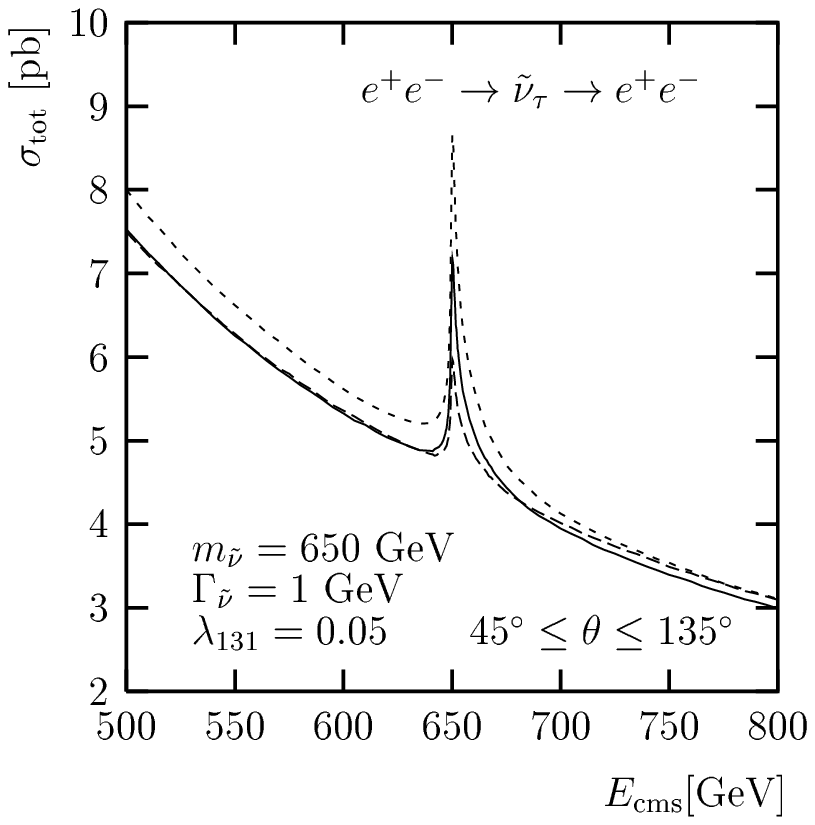}}
\put(22,45){a) $e^+ e^-\to \tilde{\chi}^0_1 \tilde{\chi}^0_2\to 
2 \tilde{\chi}^0_1 e^+ e^-$}
\put(117,45){b)}
\put(65,-8){\small $ P_{e^-}$}
\put(12,40){\small $ P_{e^+}$}
\put(53,4){\footnotesize 10\%}
\put(50,8){\small 8\%}
\put(46,12){\small 4\%}
\put(43,16){\small 0\%}
\put(37,21){\small $-4\%$}
\put(32,26){\small $-8\%$}
\put(22,36){\small $-12\%$}
\end{picture}\par\vspace{.7cm}
\caption{\it a) Contour lines of the forward--backward
asymmetry of the decay electron
$A_{FB}$/\% of
$e^+ e^-\to\tilde{\chi}^0_1\tilde{\chi}^0_2, \tilde{\chi}^0_2\to
\tilde{\chi}^0_1 e^+ e^-$
at $\sqrt{s}=(m_{\tilde{\chi}^0_1}+m_{\tilde{\chi}^0_2})+30$~GeV
in the reference scenario as a function of a) $P_{e^-}$ and $P_{e^+}$ for fixed
$m_{\tilde{e}_L}=176$~GeV, $m_{\tilde{e}_R}=132$~GeV;
b) Sneutrino production in R--parity violating model:
Resonance production of
$e^+ e^-\to \tilde{\nu}$ interfering with Bhabha scattering
for different configurations of
beam polarization: unpolarized case (solid),
$P_{e^-}=-80\%$ and $P_{e^+}=+60\%$ (hatched),
$P_{e^-}=-80\%$ and $P_{e^+}=-60\%$ (dotted) \cite{Spiesberger}. }
\label{fig:th}
\end{figure}
\end{center}

Analogues to the chargino case and the indirect constraining of the 
sneutrino mass it is possible to constrain the selectron masses indirectly
via the analysis of forward--backward asymmetries of neutralino decay leptons
\ci{Gudi_Char_Neut}. Since neutralinos are Majorana fermions
the neutralino production is exactly
forward--backward symmetric if CP is conserved. However, 
due to spin correlations between production and decay, non vanishing 
asymmetries $A_{FB}$ of the decay electron can occur \ci{GMP2}.
Beam polarization enlarges these asymmetries by about a factor 3 if
both beams are simultaneously polarized. With (85,60), e.g., 
in the reactions $e^+ e^-\to \tilde{\chi}^0_1\tilde{\chi}^0_2$, 
$\tilde{\chi}^0_2\to \tilde{\chi}^0_1 e^+ e^-$ the asymmetry is
about $4\%$ in the case of only polarized electrons but up to
$13\%$ if both beams are polarized, Fig.~\ref{fig:th}a. 
Since these asymmetries are very 
sensitive to the mass of the exchanged selectrons 
it is possible to constrain the slepton masses indirectly. 

The MSSM contains four neutralinos. One additional
Higgs singlet yields the (M+1)SSM with 5 neutralinos. 
Superstring--inspired E$_6$--models with additional  
neutral gauge bosons or Higgs singlets 
have a spectrum of six or more neutralinos. 
In certain regions of the parameter space, where the
lightest neutralino is singlino--like, the same
mass spectra of the light neutralinos are possible in the MSSM, 
(M+1)MSSM and E$_6$.
Since beam polarization is sensitive to the different couplings, 
it is a powerful tool for distinguishing 
between these models  \ci{Hesselbach}.

{\it R--parity violating SUSY:}
In R--parity violating SUSY, processes can occur which prefer the 
extraordinary (LL) or (RR) polarization configurations.
An interesting example is $e^+ e^-\to \tilde{\nu} \to e^+ e^-$,
Fig.~\ref{fig:th}b. 
The main background to this process is Bhabha scattering. 
Polarizing both electrons and positrons can strongly enhance the signal.
A study \ci{Spiesberger} was made for $m_{\tilde{\nu}}=650$~GeV,
$\Gamma_{\tilde{\nu}}=1$~GeV, with an angle cut of 
$45^{0}\le \Theta\le 135^{0}$ and a lepton--number violating coupling 
$\lambda_{131}=0.05$ in the R--parity violating Langrangian
${\cal L}_{\not R}\sim \sum_{i,j,k}\lambda_{ijk} L_i L_j E_k$. 
Here $L_{i,j}$ 
denotes the left--handed lepton and squark superfield and $E_k$ the
corresponding right--handed field \ci{Spiesberger}.
The cross section $\sigma(e^+ e^-\to e^+ e^-)$ including
$\sigma( e^+ e^-\to \tilde{\nu}\to e^+ e^-)$ gives i) 7.17 pb
(including Bhabha--background of 4.50 pb) for the unpolarized case, ii) 7.32 pb
(including Bhabha--background of 4.63 pb) for $P_{e^-}=-80\%$ and
iii) 8.66 pb (including Bhabha--background of 4.69 pb) for
$P_{e^-}=-80\%$, $P_{e^+}=-60\%$. This means that the
electron polarization
enhances the signal only slightly by about 2\%, whereas the simultaneous
polarization of both beams with $(-80,-60)$
produces a further increase by about 20\%.
This configuration of beam polarizations, which strongly suppresses pure SM 
processes, allows one to perform fast diagnostics 
for this R--parity violating process. For example the process
$e^+ e^-\to Z' $ could lead to a similar resonance peak, but with
different polarization dependence. In the latter case only
the `normal' configurations $LR$ and $RL$ play a role and its rates will be 
strongly suppressed by $LL$.

\begin{figure}
\begin{center}
\hspace{-1.3cm}
\begin{picture}(90,70)
\put(-60,0){\includegraphics{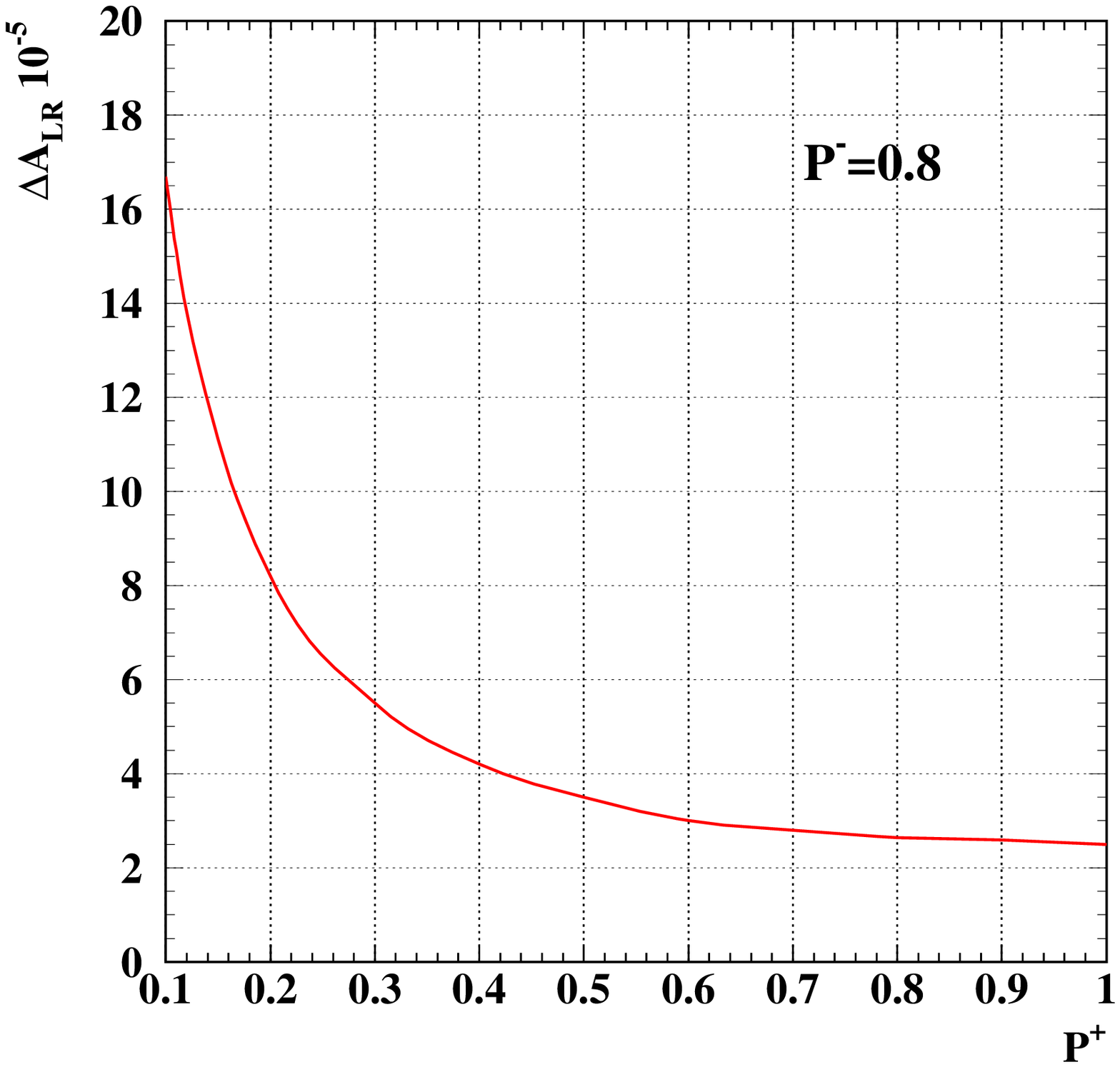}}
\put(27,90){$e^+ e^- \to Z \to \ell \bar{\ell}$}
\end{picture}
\par\vspace{+.5cm}
\caption{ Test of Electroweak Theory:
The statistical error on the left--right asymmetry
$A_{LR}$ of $e^+ e^-\to Z\to \ell \bar{\ell}$ at GigaZ
as a function of the positron polarization $P(e^+)$
for fixed electron polarization $P_{e^-}=\pm 80\%$
\cite{Moenig}.\la{fig_ew1}}
\end{center}
\end{figure}

\vspace{-1cm}
\section{Conclusion}
\vspace{-.3cm}
The clean initial state of $e^+ e^-$ collisions in a linear 
collider is 
ideally suited for the search for new physics, and the determination of both 
Standard Model and New Physics couplings with high precision.
Polarization effects will play a crucial role in these processes.
We have shown that simultaneous polarization of both beams 
can 
significantly expand the accessible physics opportunities. 
A recurring theme in this paper is that the simultaneous polarization of both 
electrons and positrons can be used to determine
 quantum numbers of new particles,
increase rates, suppress background, raise the effective polarization,
reduce the error in determining the effective polarization, distinguish 
between 
competing interaction mechanisms, and expand the range of measurable 
experimental 
observables. These virtues help to provide us with unique new insights into 
Higgs, 
Electroweak, QCD, Alternative Theories and SUSY. In particular it allows 
to enlight the structure of the underlying model.

\vspace{.3cm}
The author  would like to thank H. Steiner for  collaboration on various issues
of the work presented here, and M. Battaglia for many helpful discussions.
GMP was partially supported by the DPF/Snowmass Travel Fellowship from the
Division of Particles and Fields of the American
Physical Society, and of the Snowmass 2001 Organizing Committee.

\end{document}